\documentstyle[epsf,aps,preprint,rotate]{revtex} 
\tighten 
\begin{document}
\preprint{\small OUTP-96-16P}
\draft
\title{Nucleosynthesis Bounds on a Time--varying Cosmological ``Constant''}
\author{\large Michael Birkel and Subir Sarkar \vspace{1cm}} 
\address{Theoretical Physics, University of Oxford, \\ 
          1 Keble Road, Oxford OX1 3NP \vspace{1cm}}
\date{\sf astro-ph/9605055, revised July 1996}
\maketitle
\begin{abstract}
We constrain proposed phenomenological models for a vacuum energy
which decays with the expansion of the universe from considerations of
standard big bang nucleosynthesis. Several such models which attempt
to solve the cosmological age problem are disfavoured or even
ruled out.
\end{abstract}
\vspace{1cm}
\pacs{26.35, 98.80.Cq, 98.80.Ft 
\vfill Published in {\sl Astroparticle Physics} 6 (1997) 197
}

\widetext

\section{Introduction}

Several recent cosmological observations have revived the debate on
the possibility of a cosmological constant $\Lambda$
\cite{reviews}. The most discussed problem in this connection is the
possible conflict between the age of the universe in the standard
Friedmann-Robertson-Walker (FRW) cosmology, as inferred from recent
measurements of the Hubble constant $H_0$\cite{h0}, versus the age of
the oldest stars in globular clusters \cite{age}. Constraints from the
formation of large-scale structure in the universe together with
dynamical estimates of its matter content, $\Omega_0\approx0.2$
\cite{omega}, have also inspired suggestions of a non-vanishing
cosmological constant, in order to make up
$\Omega_0+\Lambda_0/3H_0^2=1$ \cite{structure}. (Such a low value of
$\Omega_0$ may also be required to reconcile the observed high
nucleonic content of clusters of galaxies, revealed through their
X-ray emission, with the upper limit on the nucleon density from big
bang nucleosynthesis (BBN) \cite{cluster}.) These arguments limit the
present value of the cosmological constant and the associated vacuum
energy density, $\rho_{\rm v}\equiv\Lambda\,M_{\rm P}^2/8\pi$, to be
at most $\sim(3\times10^{-12}$~GeV$)^4$, whereas quantum field
theories and the inflationary paradigm generally demand a value for
the vacuum energy before the start of the FRW cosmological evolution,
which is larger by at least 100 orders of magnitude
\cite{inflation}. While there is, as yet, no theoretical understanding
of this problem \cite{lambda}, a phenomenological approach is to
invoke a {\em time-varying} cosmological ``constant''. In such a
scenario, the vacuum energy density decays from its original high
value during inflation to the small value allowed for at the present
epoch. Many such models have been proposed in recent years
\cite{freese}-\cite{berman}.

An important constraint on models in which the vacuum energy decays
into radiation is provided by nucleosynthesis arguments, as was noted
by Freese {\it et al.} \cite{freese}. Since their analysis, however,
there have been several developments in this area \cite{bbnrev}. Input
parameters such as nuclear cross-sections, the neutron lifetime and
the number of light neutrinos are now better known, while the
theoretical calculation of elemental abundances has become more
accurate. Moreover there have been major developments on the
observational front, requiring revision of the abundance bounds
adopted in earlier work. In this paper we correct and update this
nucleosynthesis constraint and test the viability of phenomenological
proposals for a decaying vacuum energy.

\section{Bounds from Primordial Nucleosynthesis}

In considering the possible form of the time variation of
$\Lambda(t)$, it is reasonable to assume that the vacuum energy is
released as radiation, the energy density of which redshifts with the
expansion scale factor as $R^{-4}$. (If the decaying vacuum energy
creates massive particles instead, the present day energy density in
matter will increase, exacerbating the cosmological age problem
referred to above.) Hence Freese {\it et al.}  \cite{freese}
considered that the vacuum energy density decreases proportionally as
the energy density of radiation which they took to include $e^+e^-$
pairs and $N_\nu=3,4,5$ species of neutrinos in addition to photons,
i.e. $\rho_{\rm r}=\rho_\gamma+\rho_{e}+\rho_\nu\equiv\case{\pi^2}{30}g_{\rm
eff}T^4$. To conveniently track the changing vacuum energy density
they introduced the parameter
\begin{equation} \label{xfreese}
 x \equiv \frac{\rho_{\rm v}}{\rho_{\rm r} + \rho_{\rm v}}\ ,
\end{equation}
which they supposed remains constant with the expansion, even as
$\rho_{\rm v}$ decays increasing $\rho_{\rm r}$. Although apparently
plausible, this assumption is in fact physically inconsistent during
the nucleosynthesis era for the following reasons. Since the weak
interactions freeze out at a few MeV, the neutrinos generated by the
decaying vacuum energy cannot be thermalized at lower
temperatures. Further, $e^+e^-$ pairs turn non-relativistic and
annihilate soon thereafter. Consequently, to have $\rho_{\rm v}$
decrease proportionally to $T^4$ as desired for phenomenological
reasons, we must assume that the vacuum energy decays into photons
{\em alone} and redefine
\begin{equation} \label{xdef}
 x \equiv \frac{\rho_{\rm v}}{\rho_{\gamma} + \rho_{\rm v}}\ .
\end{equation}
This does remain constant throughout the BBN epoch, since the photons
produced by the decaying vacuum energy do get thermalized (within a
Hubble time) down to a temperature of about $10$~keV.

To quantitatively study the effects on nucleosynthesis, two important
changes have to be made to the BBN computer code
\cite{bbncode}. First, we add a term to the Friedmann equation to
account for the vacuum energy density:
\begin{equation}
 H^2 = \frac{8\pi}{3M_{\rm P}^2} (\rho_{\rm r} + \rho_{\rm v})\ .
\end{equation}
(We have neglected the energy density of non-relativistic matter,
$\rho_{\rm m}$, as is appropriate in the radiation-dominated era.)
Secondly, the equation of local energy conservation now has a term
$\dot{\rho}_{\rm v}$ describing the production of energy by the
decaying vacuum energy:
\begin{equation}
 \dot{\rho}_{\rm v} + \dot{\rho} + 4\frac{\dot{R}}{R} \rho = 0\ . 
\end{equation}
Freese {\it et al.} \cite{freese} had stated that this leads to the modified
time-temperature relationship
\begin{equation}
 T(t) \propto [g_{\rm eff} (1-x)]^{-1/4} t^{-1/2}\ ;
\end{equation}
this, however, obtains only if the effects of $e^+e^-$ annihilation as
well as the relative decrease of the neutrino temperature due to
vacuum decay are neglected. (Because of the entropy generated by
vacuum energy decay, the temperature of the background neutrinos (as
well as the nucleon-to-photon ratio $\eta$) continually decreases in
ratio to the photon temperature even at $T{\ll}m_e$ and drops below
its standard asymptotic value of $T_\nu=(4/11)^{1/3}T$.) We discuss
this in the Appendix and implement these effects in the Wagoner code
as updated by Kawano \cite{bbncode}. We assume 3 light neutrino
species, use the latest value of the neutron lifetime
$\tau_{n}=887\pm2$~s \cite{pdg}, and incorporate small corrections to
the helium abundance as reviewed in ref.\cite{bbnrev}.

\begin{figure}[t]
\rotate[r]{\epsfysize12.5cm\epsfbox{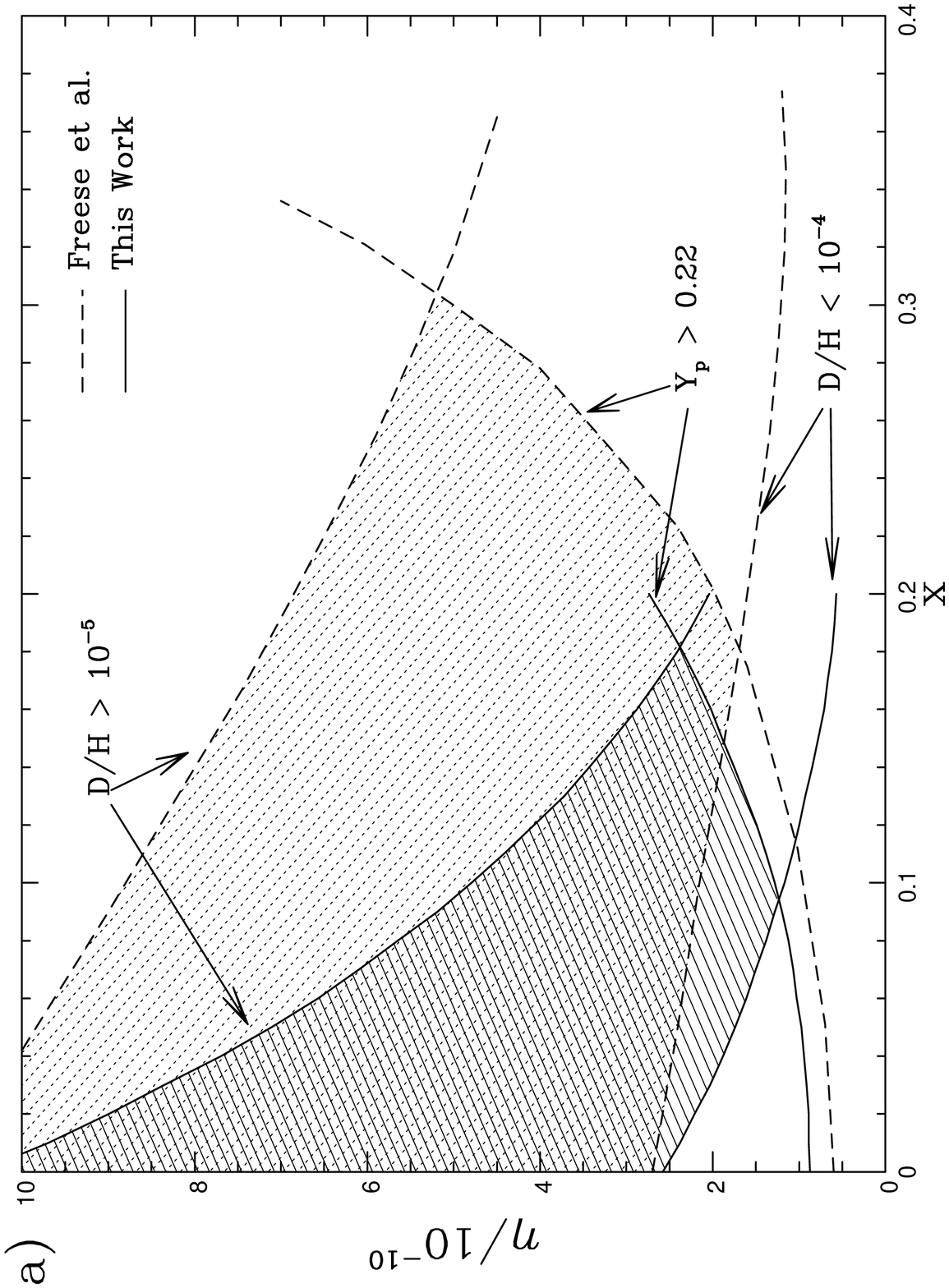}}
\rotate[r]{\epsfysize12.5cm\epsfbox{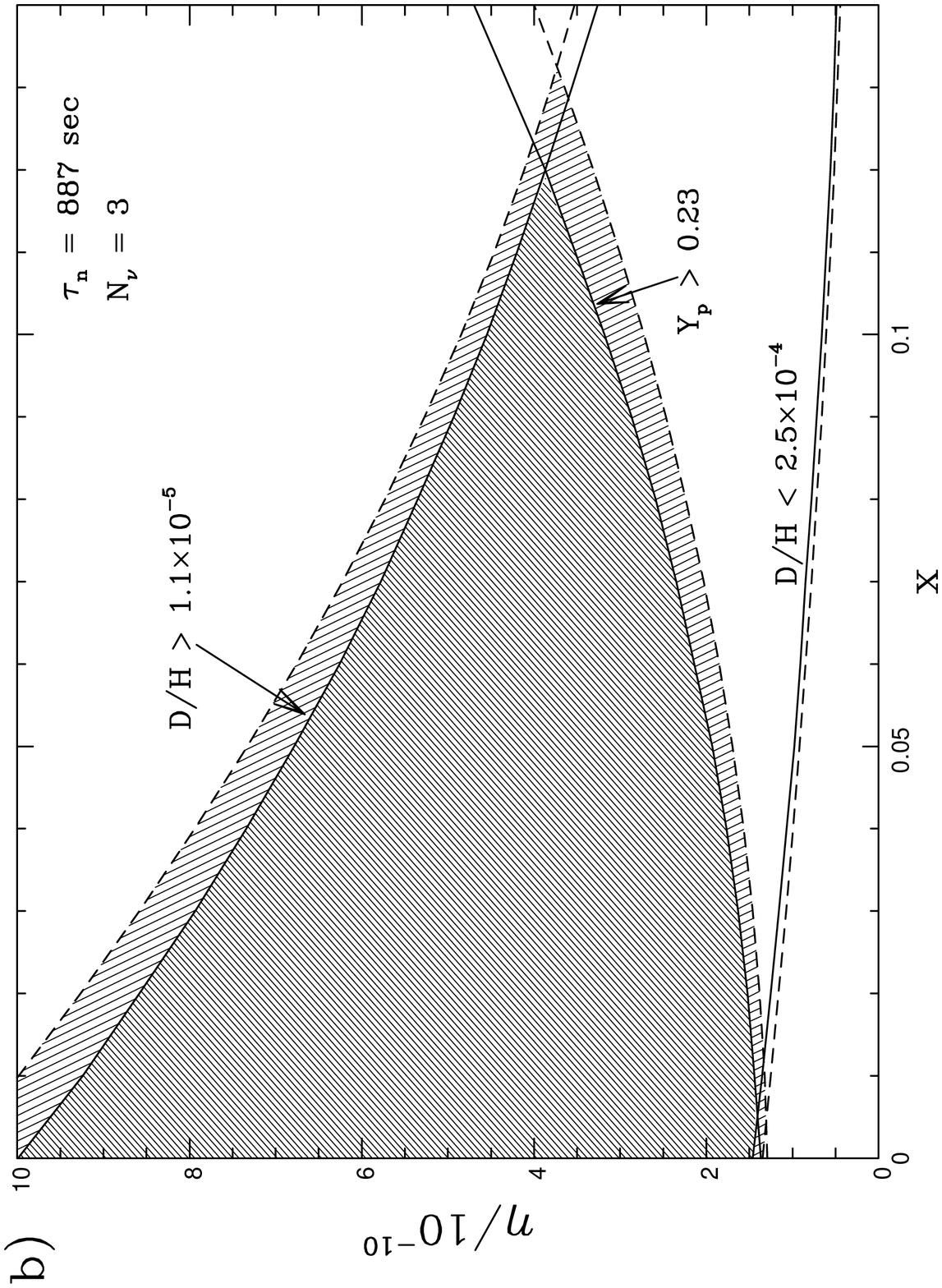}}
\caption{Bounds on the decaying vacuum energy parameter
$x\equiv\rho_{\rm v}/(\rho_\gamma+\rho_{\rm v})$ as a function of the
nucleon-to-photon ratio $\eta$ (evaluated at $T=10^8$~K). The shaded
regions are allowed. In panel (a) we compare previous results (dashed
lines) with those obtained using the improved nucleosynthesis code,
adopting the same bounds on the elemental abundances. In panel (b) we
adopt updated observational bounds and also show (dashed lines) the
``$95\%$'' C.L. limits corresponding to the known uncertainties in
input parameters.}
\label{fig1}
\end{figure}

The overall effect of a vacuum energy decaying to produce entropy is a
{\em decrease} in the synthesized abundance of $^4$He. This is because
the effect of the increased expansion rate, which causes earlier
neutron-proton freeze-out, is sub-dominant relative to the effect of
the dilution of $\eta$, which delays the onset of nuclear
reactions. Consequently, although the $n/p$ ratio at freeze-out is
higher, it decreases by $\beta$-decay to a smaller value before the
neutrons can be synthesized in light nuclei.\footnote{Note that if the
decays do create (electron) neutrinos as assumed by Freese {\it et
al.}  \cite{freese}, then there will be further {\em direct} effects
on neutron-proton interconversions; such effects, however, were not
considered by these authors.} Thus the magnitude of the decaying
vacuum energy during nucleosynthesis can be bounded by requiring that
the abundances of the synthesized elements be within observational
limits. The relevant bounds considered in earlier work \cite{freese}
were
\begin{equation} \label{abundfreese}
 0.22 \leq Y_{\rm p}(^4{\rm He}) \leq 0.26 , \qquad
 10^{-5} \leq ({\rm D/H})_{\rm p} \leq 10^{-4} ,
\end{equation}
for the primordial mass fraction of helium and the primordial
abundance by number of deuterium, respectively. We adopt these bounds
as well in order to illustrate the difference in results due to our
use of an improved code and compare the constraints obtained on $x$ as
a function of $\eta$ in Figure~\ref{fig1}(a), having rescaled the
previous result \cite{freese} to correspond to our redefinition
(\ref{xdef}) of $x$, taking $\rho_{\rm r}=\case{43}{8}\rho_\gamma$,
for $N_\nu=3$. (Note that the constraints corresponding to the upper
bound on $Y_{\rm p}$ are out of range, to the left of the y-axis). The
constraint quoted earlier \cite{freese} corresponds to a bound on the
vacuum energy which is less restrictive than our result by a factor of
about 2.

Next we impose the updated bounds on the element abundances inferred
from recent observations of helium in metal-poor extragalactic HII
regions \cite{izotov} and of deuterium in high redshift clouds along
the line of sight to quasars \cite{hogan}:
\begin{equation} \label{abund}
 0.23 \leq Y_{\rm P}(^4{\rm He}) \leq 0.25 , \qquad
 1.1 \times 10^{-5} \lesssim ({\rm D/H})_{\rm p} \lesssim 2.5 \times 10^{-4} .
\end{equation}
(We do not consider the bound on $^7{\rm Li}$ because it does not
provide a useful constraint.) We emphasize that these are {\em
conservative} bounds based upon consideration of a variety of data
which are critically discussed in ref.\cite{bbnrev}. Because of the
uncertainties in the input nuclear cross-sections and the neutron
lifetime, as well as residual numerical errors in the computer code,
the yield of an element for some specified value of $\eta$ and $x$ is
not uniquely determined but has a spread which is about $\pm0.5\%$ for
helium and as much as $\pm50\%$ for deuterium. Using a Monte Carlo
method in which the BBN code is run many times while the input
parameters are sampled at random from their known distributions, the
``$95\%$'' C.L. bound can be identified as the value at which $5\%$ of
the runs pass the imposed observational bounds \cite{kernan}. On the
basis of these results we find a maximum value of
\begin{equation} \label{xlim}
 x_{\rm max} = 0.13
\end{equation}
for $\eta\simeq3.7\times10^{-10}$ as shown in
Figure~\ref{fig1}(b). This corresponds to the bound
\begin{equation}
 \rho_{\rm v} < 4.5 \times 10^{-12}\ {\rm GeV}^4\ ,
\end{equation}
evaluated at a temperature $T=2.6$~MeV. This is about 3 times more
stringent than the often quoted constraint calculated by Freese {\it
et al.} \cite{freese}. For comparison, a {\em constant} vacuum energy
is constrained by nucleosynthesis to be $\rho_{\rm
v}<4.4\times10^{-17}$ GeV$^4$ for $\eta$ in the range
$10^{-11}-10^{-9}$. (Although the bound on a time-varying cosmological
``constant'' appears weaker, it should be noted that it would decay to
a much smaller value by the present epoch than one which is truly
constant.)

\section{Critique of Decaying--$\Lambda$ Models}

Many of the proposed models in the literature assume that
\begin{equation} \label{timevar}
 \Lambda \propto R^{-m}\ ,
\end{equation}
where $1\leq{m}\leq3$ \cite{irrelev}. Since the associated vacuum
energy density $\rho_{\rm v}$ then decreases more slowly with time
than the energy density of radiation, even a $\rho_{\rm v}$ having its
maximum allowed value today would have been negligible at the epoch of
nucleosynthesis. However, it would be physically more consistent to
normalize the vacuum energy density at some high energy scale instead
of at the present epoch. For example Gasperini \cite{gasper}
interprets the cosmological constant as a measure of the Hawking
temperature of the De Sitter vacuum, $T_{\rm
v}=(\Lambda/12\pi^2)^{1/2}$; he assumes that the vacuum temperature
equals the radiation temperature close to the Planck era, but that the
subsequent expansion causes $T_{\rm v}$ to decay faster than $T$ so as
to be consistent with the upper bound on the cosmological constant
today. However, the value of $\rho_{\rm v}$ during nucleosynthesis
would still be substantial; in order to obey the constraint
(\ref{xlim}), the exponent in eq.(\ref{timevar}) is now required to be
\begin{equation}
 m > 3.5 .
\end{equation}
Thus Gasperini's suggestion is inconsistent with many models
\cite{irrelev} which assume $m$ to be less than 3. (Note that it does
not matter that the constraint (\ref{xlim}) was derived assuming $m=4$
since it would be even more stringent for smaller $m$.)

Some other models \cite{contrived} have too many free parameters
and/or are not specified sufficiently explicitly to be confronted with
BBN. Specific models motivated by physical considerations
\cite{limtrod,wett,moff,berman} can however be constrained, either by
the bound (\ref{xlim}), or by the related bounds on the expansion rate
or the values of fundamental constants during BBN. In some cases the
authors have already considered such constraints but commented that
the associated uncertainties preclude a definitive test. We reexamine
the question taking into account all such sources of error and
imposing {\em conservative} constraints from BBN \cite{bbnrev}.

Lima and collaborators \cite{limtrod} consider nonsingular
deflationary cosmology models with a decaying vacuum energy density.
The cosmological history starts with the decay of a De Sitter vacuum
and evolves smoothly to a quasi--FRW stage at late times. Irrespective
of the spatial curvature, the models are characterized by the time
scale $H_{\rm I}^{-1}$ which determines the initial temperature of the
universe as well as the largest value of the vacuum energy
density. The present matter and radiation content of the universe is
generated by the slow decay of the vacuum energy density; for its time
dependence, the authors use the phenomenological ansatz
\begin{equation}
 \rho_{\rm v} = \frac{\Lambda M_{\rm P}^2}{8 \pi} 
   = \beta \rho_{\rm tot} 
     \left(1 + \frac{1-\beta}{\beta} \frac{H}{H_{\rm I}}\right)\ ,
\end{equation}
where $\rho_{\rm tot}=\rho_{\rm v}+\rho_{\rm m}+\rho_{\rm r}$ and
$\beta$ is a dimensionless parameter of order unity. This
automatically generates a primordial inflationary scenario at the time
$H_{\rm I}^{-1}$. The model thus represents a generalization of
ref.\cite{freese} since it does not fix the spatial curvature to be
zero and introduces a time dependence in the parameter
\begin{equation} \label{limmm}
  x' \equiv \frac{\rho_{\rm v}}{( \rho_{\rm v} + \rho_{\rm m} + 
                 \rho_{\rm r})} = \beta + 
                 (1 - \beta) \frac{H}{H_{\rm I}}\ .
\end{equation}
At late times, when the term $H/H_{\rm I}$ is negligible, we recover
the model by Freese {\it et al.} \cite{freese} and find that
$x'\simeq\beta$. In the physically preferred case, the deflationary
process begins at the Planck time, $H_{\rm I}^{-1}{\sim}M_{\rm
P}^{-1}$, while the latest epoch at which it can occur is electroweak
symmetry breaking (to permit the subsequent generation of a baryon
asymmetry), i.e. $H_{\rm I}^{-1}{\lesssim}M_{W}^{-1}$, so $H/H_{\rm
I}$ will necessarily be negligible during nucleosynthesis. Thus the
nucleosynthesis constraint (\ref{xlim}) translates into the
requirement
\begin{equation}
  \beta < 0.13 \ ;
\end{equation}
the inequality is reinforced by the fact that $x$ as defined by us
(eq.(\ref{xdef})) is always bigger than $x'$. This conflicts with the
lower limit on $\beta$ \cite{limtrod} following from the requirement
of a sufficiently long age for the universe:
\begin{equation}
  \beta \geq 0.21 \ ,
\end{equation}
thus {\em excluding} this model \cite{limtrod}.

The model proposed by Wetterich \cite{wett} involves a scalar field
$\phi$ with the potential
\begin{equation}
 V \propto \exp \left(-\frac{4\sqrt{\pi}}{M_{\rm Pl}}a\phi\right) ,
\end{equation}
where $a>0$ is a parameter. Such exponential potentials may well arise
in models of unification with gravity such as Kaluza-Klein theories,
supergravity theories or string theories. The scalar field couples to
gravity and its energy-momentum tensor induces a time-dependent
cosmological ``constant''. The model further contains a coupling of
the scalar field to matter $\propto\,\beta\rho$, where $\beta$ is
another model parameter which may, in principle, have different values
during the matter-dominated and radiation-dominated eras. Wetterich
assumes that $\beta_{\rm r}$ is zero and finds that $\beta_{\rm
m}\leq0$ from considerations of the stability of condensed objects
with a static energy density. After some critical transition time
$t_{\rm tr}$, the cosmological ``constant'' adjusts itself dynamically
to become proportional to the energy density $\rho$ (of radiation plus
matter), and all energy densities ($\rho$, $V$, $\dot\phi^{2}/2$)
decay $\propto\,t^{-2}$. Wetterich considers two possible scenarios
for the initial epoch before this transition. In the first, the
universe is initially $\rho$-dominated and the scalar field
contributions do not influence the time evolution of the
scale factor. In the second, the universe starts off being
$\phi$-dominated; this would lead to an exponential increase of the
scale factor as in vacuum energy driven inflation, although without
any subsequent entropy production.\footnote{This is unappealing since
the horizon and flatness problems of the universe {\em cannot} be
solved if there is no such entropy production \cite{wutuwe}.} As an
alternative to the above possibilities, a purely scalar-gravity system
is also considered; this necessitates additional fine-tuning to evade
the severe experimental bounds on such a theory. Thus only the first
case, viz. initial $\rho$-domination, appears to be well motivated.

The parameter $\beta$ can now be constrained as follows. In the
matter-dominated era for distances much smaller than $H^{-1}$, the
coupled system of gravity and small scalar fluctuations around the
cosmological value $\phi(t)$ is identical to the standard Brans-Dicke
(BD) theory. Experimental bounds on the Brans-Dicke parameter $\omega$
then constrain $\beta_{\rm m}$ to be small \cite{wett}:
\begin{equation}
 |\beta_{\rm m}| < 0.016 \ .
\end{equation}
If in fact $\beta_{\rm m}$ is identically zero, then the contribution
of the scalar field today is
\begin{equation}
 \Omega_{\phi_0} = \frac{3}{2 a^2} = 1 - \Omega_{{\rm m}_0}\ .
\end{equation}
Thus $\Omega_{\phi_0}>0$ corresponds to $a^2>3/2$. We can obtain
a stronger bound on $a$ if the transition time $t_{\rm tr}$ is smaller
than the BBN epoch so that the increased total energy density during
BBN due to the scalar contributions speeds up the expansion as
$t\to{t'}=\xi^{-1}t$. The recently revised \cite{bbnrev} upper limit
on such a speed-up factor $\xi$ is
\begin{equation} \label{xi}
 \xi < 1.12\ , 
\end{equation}
corresponding to the conservative abundance bounds in
eq.(\ref{abund}).\footnote{Wetterich \cite{wett} takes this limit to
be $\xi^2-1<0.1$, which then requires $a^2>22$.} This translates into
the lower bound
\begin{equation}
 a^2 > 9.9 \ ,
\end{equation}
where we have used the relation $\xi^2-1=2/|a^2 - 2|$. In turn this
requires
\begin{equation}
 \Omega_{{\rm m}_0} > 0.85
\end{equation}
at the present epoch, i.e. the present cosmological constant
contribution must be quite small. A possible escape route is to assume
$\beta_{\rm m}\neq0$, but this leads to a time dependence of the
nucleon mass and, possibly, other fundamental constants (which can
also be constrained by nucleosynthesis if the model is made explicit
enough). Even so the modification to the age of the universe,
$t_0=\case{2}{3}H_0^{-1}(1-\case{\beta}{a})$, is negligibly small,
removing the major motivation for the scenario. This conclusion may
however be evaded by assuming that the parameter $a$ is not constant
but varies with the curvature scalar or the value of the scalar
field. Another escape route would be to assume that $t_{\rm tr}$ is
larger than the BBN epoch, so the expansion rate at this time is
unaltered. It is evident that the requirement of successful
nucleosynthesis imposes interesting constraints on the possibilities
considered in ref.\cite{wett}.

Moffat \cite{moff} considers a model in the context of a BD theory
where the non-minimal coupling of a scalar field to gravity allows for
a negative pressure associated with the kinetic energy of the
field. The cosmological ``constant'' is assumed to be space-time
dependent and is equated to the kinetic energy in the BD theory. As a
result the evolutionary equation for the scalar field in a FRW
universe produces an attractor mechanism which drives $\Lambda$
towards a minimum of the potential and leads to a small constant value
for $\Lambda$ today. In this model Moffat finds a solution to the
problem of the age of the universe with $\Omega_{\rm v}$ several times
larger than $\Omega_{\rm m}$ at present. However, the constraint
(\ref{xi}) on the speed-up factor $\xi$ during nucleosynthesis
requires $\Omega_{\rm v}\lesssim0.1$, thus removing the motivation for
this model.

Berman \cite{berman} assumes $\Lambda\propto t^{-2}$ in a BD theory,
leading to a possible time variation in the gravitational constant
$G_{\rm N}$. If, in fact, $G_{\rm N}$ is constant, the vacuum energy
density equals the energy density in radiation, so that $x$ as defined
in eq.(\ref{xfreese}) is 0.5 and thus in conflict with the
nucleosynthesis constraint (\ref{xlim}). On the other hand, if $G_{\rm
N}$ varies with time, it must do so in this model as $G_{\rm N}\propto
t^{-1}$ during matter-domination and as $G_{\rm N}\propto t^2$ during
radiation-domination \cite{berman}. Such a strong time variation is
{\em ruled out} given that the value of $G_{\rm N}$ during
nucleosynthesis is required to be within about $25\%$ of its present
value \cite{timedepG}.

\section{Conclusions}

We have examined proposed models of a time-varying cosmological
``constant'' which can significantly increase the expansion age of the
universe, even for a large Hubble parameter today, thus evading the
potential age problem. However, these models also imply a deviation
from the standard expansion history during primordial
nucleosynthesis. We find that even conservative observational limits
on the abundances of the synthesized light elements imply severe
constraints on such models, demonstrating the power of this
cosmological probe.

\acknowledgments{M.B. gratefully acknowledges financial support from
the Fellowship HSP II/AUFE of the German Academic Exchange Service
(DAAD). S.S. acknowledges a PPARC Advanced Fellowship and support from
the EC Theoretical Astroparticle Network.}

\appendix

\section*{}

\begin{figure}[t]
\rotate[r]{\hskip2cm\epsfysize12.5cm\epsfbox{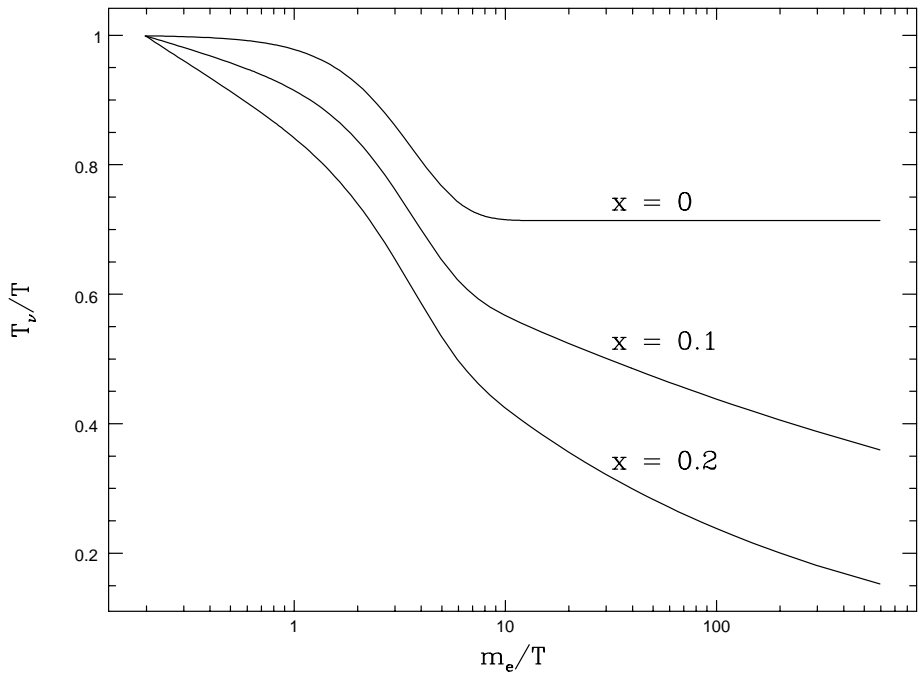}}
\caption{The evolution of the neutrino temperature after decoupling
from kinetic equilibrium for various assumed values of the decaying
vacuum energy parameter $x\equiv\rho_{\rm v}/(\rho_\gamma+\rho_{\rm
v})$.}
\label{fig2}
\end{figure}

As mentioned earlier, the ratio of neutrino to photon temperature
continues to decrease even after $e^+e^-$ annihilation. To study this,
we follow the standard procedure \cite{weinberg} of combining the
equations describing the change in entropy and energy conservation to
obtain
\begin{equation}
 S(T) - S(T_i) = \frac{4 \pi^2 x}{15 (x-1)} 
                 \int_{T_i}^{T} {\rm d} \tilde{T}\ R^3 \tilde{T^2} \ ,
\end{equation}
where $\rho_{\rm v}=x\rho_{\gamma}/(1-x)$ and
$\rho_{\gamma}=\pi^2T^4/15$. Inserting the expressions for
$S(T)$ and $S(T_i)$, $S(T)=R^3(\rho + p)/T$, we can rewrite
the above equation to obtain
\begin{equation} \label{tnu}
 T^3_\nu = T^3 \left(\frac{4}{11}\right) 
            \left[1 + \frac{45}{2 \pi^4} \left( I^{21} +
             \frac{1}{3} I^{03} \right) \right] 
            + \frac{12 x}{11 (1-x) R^3} 
             \int_{T_i}^{T} {\rm d} \tilde{T}\ R^3 \tilde{T^2} \ , 
\end{equation}
\begin{equation}
 I^{mn} \equiv \int_{m_e/T}^{\infty}  {\rm d} y\ y^m 
  \left[y^2 - \left(\frac{m_e}{T}\right)^2\right]^{n/2} 
  \frac{1}{{\rm e}^y + 1}\ .
\end{equation}
The first term in the sum above tracks the standard behaviour of the
neutrino temperature during $e^+e^-$ annihilation. The numerical
solution of this equation is shown in Figure~\ref{fig2} for several
values of the parameter $x$; note that the curves converge at the
neutrino decoupling temperature.

Freese {\it et al.} \cite{freese} did not take into account the
differing neutrino and photon temperatures or the change in the number
of degrees of freedom due to $e^+e^-$ annihilation, hence their
relation for the nucleon-to-photon ratio, $\eta=\frac{n_{\rm
N}}{n_{\gamma}}\propto T^{3x/(1-x)}$, only holds {\em after} this
epoch. Between neutrino decoupling and $e^+e^-$ annihilation, the
correct relation should be
\begin{equation}
 \eta \propto T^{12x/11(1-x)} \ .
\end{equation}
Similarly, the equation for $R(T)$ as given by Freese {\it et al.},
viz. $R\propto T^{1/(x-1)}$, which holds only at $T{\ll}m_e$, is
modified to
\begin{equation}
 R \propto T^{(11-7x)/11(x-1)} 
\end{equation}
between neutrino decoupling and $e^+ e^-$ annihilation.

\end{document}